# Wearable Internet of Things for Personalized Healthcare: Study of Trends and Latent Research

**Samiya Khan[1], Mansaf Alam[1]**

Department of Computer Science, Jamia Millia Islamia, New Delhi, Delhi
[1]samiyashaukat@yahoo.com, [1]malam2@jmi.ac.in

**Abstract: In this age of heterogeneous systems, diverse technologies are integrated to create application-specific solutions. The recent upsurge in acceptance of technologies such as cloud computing and ubiquitous Internet has cleared the path for Internet of Things (IoT). Moreover, the increasing Internet penetration with the rising use of mobile devices has inspired an era of technology that allows interfacing of physical objects and connecting them to Internet for developing applications serving a wide range of purposes. Recent developments in the area of wearable devices has led to the creation of another segment in IoT, which can be conveniently referred to as Wearable Internet of Things (WIoT). Research in this area promises to personalize healthcare in previously unimaginable ways by allowing individual tracking of wellness and health information. This chapter shall cover the different facets of Wearable Internet of Things (WIoT) and ways in which it is a key driving technology behind the concept of personalized healthcare. It shall discuss the theoretical aspects of WIoT, focusing on functionality, design and applicability. Moreover, it shall also elaborate on the role of wearable sensors, big data and cloud computing as enabling technologies for WIoT.**

## 1. Introduction

Recent past has seen a dramatic rise in the incidence of chronic, life threatening diseases. Moreover, the cost of healthcare rests on an ever-rising curve. Thus, there is an urgent need to transmute the healthcare providers' approach from hospital-centric to patient-centric. In order words, in the modern scenario, focusing health efforts on personalized disease management and individual well being, make most sense. The motivation behind the use of technology in health is aimed at providing e-health and m-health services to individuals, targeting to improve the operational efficiency of the healthcare system. Smart phones and mobile-based service provisioning are some of the most path breaking technological advancements of this era. According to a study [1], 21.2% of mobile users in India owned a smart phone in 2014 and this percentage is expected to increase to 36.2% by 2022, projecting an approximate rise of 70%. The graphical illustration is shown in Fig. 1. It is important to note that values for years 2014-2018 are actual values while the same for 2019-2022 are projections. Moreover, with slashed data usage costs, the number of India-based Internet users is expected to witness a rise of 40% by 2023 [2].

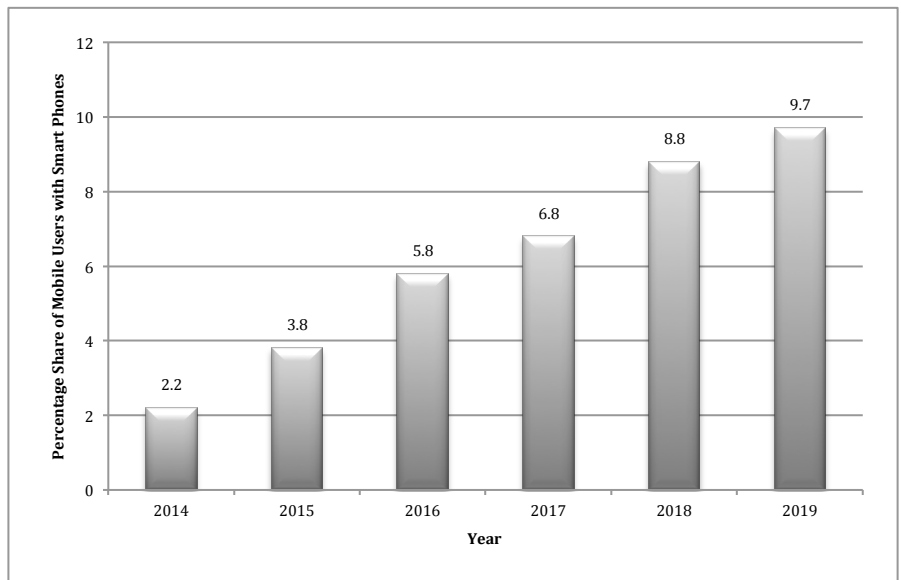

Fig. 1: Smart Phone Users' Share in India [2]

These statistics set the ground for a mobile-based personalized healthcare model, in which smart phones can be used for sensing, connectivity and establishing interaction with the individual, making it a core driving technology for mHealth [3]. With that said, mobile devices come with basic sensing facilities like motion tracking and activity recording. It is not possible for a standalone smart phone to capture the finer details and parametric assessment of an individual's health. It is for this reason that wearable devices have to be integrated with this model for expanding the sensing capabilities of a smart phone.

The wearable technology segment has grown immensely in the recent past recording a 5.3% increase in revenues for the U.S. market in 2018 [4]. Fig. 2 shows a graphical illustration of consumer revenue generated for year 2014 and 2015, with value projections for the years ranging between 2016-2019. Moreover, the user penetration of this technological segment for the American market stood at 11.8% [4]. As reported by Quid [4], majority of the people who buy wearable products use it for health, fitness and clinical reasons. Evidently, this market is expected to grow in the coming years for the obvious cause that a consumer can buy multiple wearable products and their applications are not limited to just fitness and health. Wearable technology is increasingly being purchased because of other reasons as well, such as improved connectivity and innovative value.

Wearable devices are capable of serving multiple functions, which include (1) body data collection (2) basic preprocessing (3) transient data storage, and (4) data transfer to server or mobile phone. The unique selling point of this technology is 'wearability', which allows collection of explicit data values on the basis of application-specific requirements. Fundamental advantages of using this technology include automatic diagnostic monitoring and allowance of timely interventions. On the hindsight, this technology is also confronted with significant challenges such as short communication bandwidth, limited battery life and restricted computing ability.

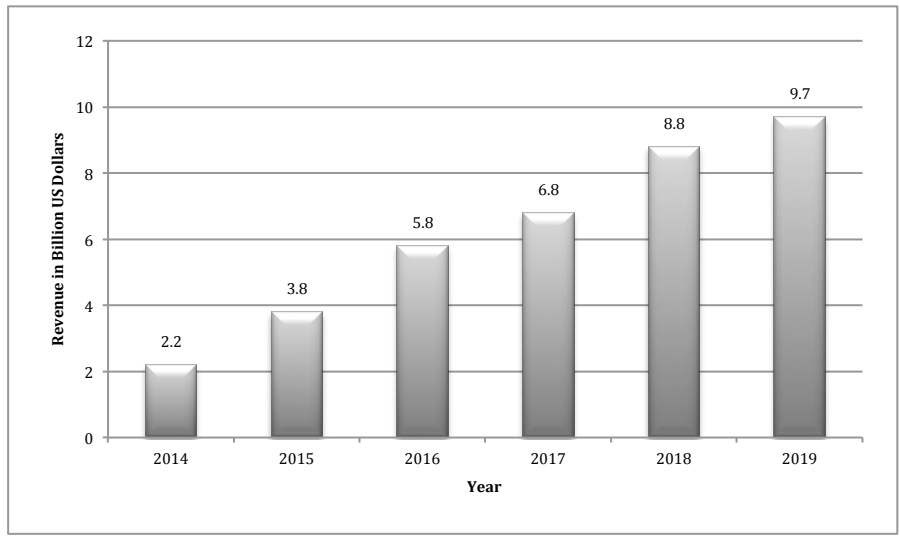

Fig. 2: Revenue Analysis of Consumer Wearable in US [4]

Internet of Things provides a robust technological framework by facilitating data collection from wearable sensors and mobile devices, and backing up the computing and storage capabilities by integrated use of state of the art technologies like cloud computing [69] and big data [70]. The rest of the chapter is organized in the following manner: Section 2 introduces the concept of Wearable Internet of Things (WIoT) and covers background concepts related to the same. There are a plethora of WIoT applications. However, this chapter limits its scope to personalized healthcare and Section 3 covers this facet in detail. Finally, Section 4 summarizes the challenges and future research prospects in the field with concluding remarks synopsized in Section 5.

## 2. Concept of Wearable Internet of Things (WIoT)

The ubiquitous nature of Internet connectivity has paved way for a new era in technology, that of Internet of Things. Simplistically, Internet of Things (IoT) can be described as a technology that allows Internet-backed connectivity between 'things' to allow data processing, analytics and visualization for development of domain-specific intelligent solutions [5]. Therefore, Internet of Things, essentially, uses technologies such as big data and cloud computing to empower its base framework. The past few decades have seen a swift transition from Internet to Internet-based services like social networks [6] and wearable web [7].

Traditionally, the concept of Internet of Things (IoT) is an extension of Internet to the real world and its entities. Therefore, its early versions depended on RFID technologies [8][9] for transforming 'things' into sensing devices. Heterogeneous sensors such as accelerometers and gyroscopes succeeded this technology. However, the use of IoT in healthcare was well established only after the development of wearable devices [10]. Thus, WIoT is majorly used in healthcare for improving diagnostics, allowing early interventions and enabling execution of remote surgeries [11-14].

The emergence and growing popularity of wearable devices has led to the creation of a new segment in Internet of Things (IoT), widely referred to as Wearable Internet of Things (WIoT). In this context, wearable devices form an intrinsic fiber for the intelligent fabric of IoT, which connects numerous near-body and on-body sensors with Internet and each other. Another significant facet of WIoT framework is connecting the end-points to medical infrastructure like physicians and hospitals so that longitudinal assessment of patient conditions can be performed irrespective of their location.

Collected data needs be pushed to the concerned healthcare provider and timely intervention or management of emergencies can be performed. Therefore, Wearable Internet of Things is an infrastructure that enables this interconnectivity and facilitates examining of human factors like behaviour, wellness and health, thereby contributing in making interventions for enhanced quality of life [15].

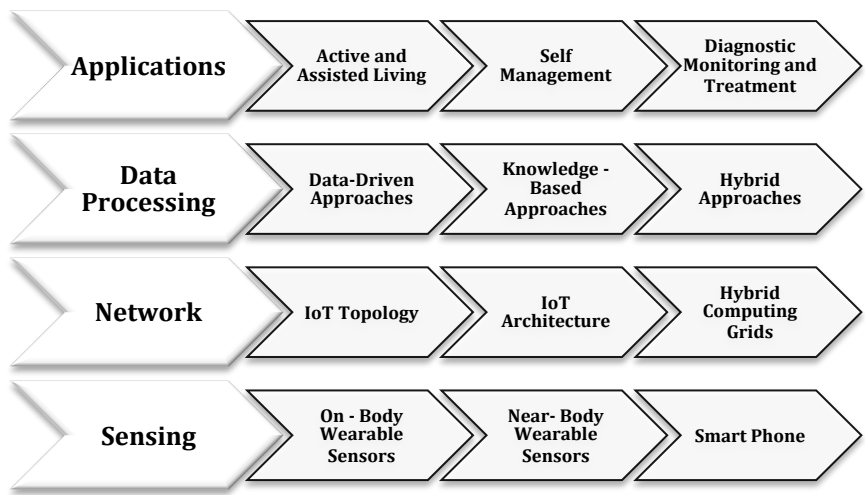

Fig. 3: Components of WIoT Architecture

Data lifecycle consists of standard phases namely collection, storage, processing and visualization [16]. In view of this, Qi et al. [17] proposed an IoT-based architecture for personalized healthcare systems. Fig. 3 illustrates the functional components of WIoT, which is inspired by the same proposed by Qi et al. [17]. However, the below-mentioned architecture focuses its attention on wearable devices and their integrated use with IoT for healthcare. The architecture divides Wearable Internet of Things (WIoT) into four functional components namely sensing, network, processing and applications. A detailed description of these functional components is provided below.

## 2.1. Sensing

The purpose of sensing layer is to efficiently acquire medical and personal information of an individual using sensing technologies. Qi et al. [17] provided a classification of available sensors. These include both near-body and on-body sensors. Details of the classification scheme are provided in Table 1. Available sensors are capable of capturing physiological information like heart activity, eye movement and blood pressure. Besides this, inertial sensors, location sensors and image sensors may be used for tracking activity.

Table 1: Classification of Sensors

| Category | Sensors | Measurements Taken |
|---|---|---|
| Inertial Sensors | Magnetic Field Sensor [18] | Higher spatial resolution's location |
| | Pressure Sensor [19] | Altitude of the object |
| | Gyroscope [20] | Angular rotational velocity |
| | Accelerometer [21] | Measures linear acceleration |
| Physiological Sensors | Galvanic Skin Response (GSR) [22] | Temperature of skin surface |
| | Electrooculography [23] | Movement of eye |
| | Spirometer [24] | Lung parameters like volume, expiration and flow rate. |
| | Electrocardiogram (ECG) [13] | Heart activity |
| | Blood pressure cuff [25] | Blood pressure |
| Image Sensors | SenseCam [26] | Pictures of daily living activities |
| Location Sensors | GPS [27] | Coordinates of outdoor location |

While near-body sensors may be helpful in gathering additional parameters for analysis, wearable body area sensors form the core component of WIoT. These sensors are responsible for capturing data from the body directly through contact or with the help of peripheral sensors for capturing behavioral patterns. Finally, these sensors preprocess data and prepare it for on-board analysis or offshore decision making. Typically, the design of a wearable device is application-dependent. However, most wearable devices can be expected to come packaged with communication capabilities, on-board power management and embedded computing device with limited storage support.

Lately, many commercial products have come into existence. Peripheral wearable sensors that can be worn on the wrist or arm like BodyMedia armband [28], are dedicated fitness monitoring wearable devices that work on minimum hardware and computing abilities to provide real time activity data of the individual. Smart watches [29] are also fitness and activity monitors, but they are generally capable of higher functionality and perform many other functions related to the smart phone as well.

Novelty in this domain largely focuses on the body-sensor interface for different kinds of parametric data collection. Some examples of such wearable sensors include Bio-Patch [30], ring sensor [31] and ECG monitor [32]. Recently, the concept of smart textiles and smart clothing have also gathered immense research attention, bringing forth the idea of embedding sensors in the textile or fabric used for creation of clothing to allow unobtrusive monitoring. These sensor fabrics are known to be extremely effective in monitoring response of human autonomous nervous system [33].

These sensors may be used in conjunction with ambient sensors such as environment monitoring sensors [34-35] and indoor localization sensors [36-37] for a comprehensive analysis of individual's intrinsic and extrinsic health. Moreover, some studies [38-41] suggest that smart phone applications' data can be used for human behavior monitoring, making smart phone, another source of data. The challenge in this domain is to develop non-invasive and cost-effective sensors for automatic data acquisition in uncontrolled surroundings like IoT-based systems and researches [42-44] are being performed in this direction.

## 2.2. Network

Sensing devices cannot function and serve their purpose in a standalone manner because of their restricted computing and storage capabilities. Moreover, limitations with communication bandwidth and power management also make them insufficient devices when alone. All the sensing devices need to be connected to the IoT infrastructure for data aggregation, storage and transmission for further analysis.

Firstly, all the sensing devices need to be configured and deployed using a standard or hybrid topology. At this level, transfering of mobile and static sensing devices to hybrid computing grids is an existing research challenge [17]. Several IoT infrastructures such as 6LoWPAN [45] have been proposed for efficient data transfer in view of the scalability and mobility issues that arise with the use of IoT systems.

Therefore, challenges at the network layer include interoperability, energy-efficiency, QoS requirements and network management of heterogeneous components. With that said, one of the core issues that need to be tackled at this level for personalized healthcare applications is security and privacy of user data. It has been proposed that Mobile Cloud Computing (MCC) paradigm can be used [15] for management of power and performance issues associated with WIoT as MCC allows cloud-based storage and analysis of data and inherently tackles the issues associated with mobility and flexibility requirements of such applications.

## 2.3. Data Processing

The medical data generated by wearable sensors and smart phones is expected to be huge. Moreover, deriving knowledge from available data is just as crucial as acquisition of data. Applications can only be developed for patient benefit after acquired data is put through intelligent algorithms to process it and facilitate action-based decisions. Early works in this domain focus on development of algorithms and computational methodologies for processing of disease-specific datasets.

Research on generalized methods for processing of medical and health data caught the attention of scientists much later with possible use of intelligent algorithms for anomaly detection, pattern recognition and decision support being investigated in later works [46]. Qi et al. [17] classifies the data processing approaches used for healthcare systems into three categories namely data-driven, knowledge-based and hybrid approaches. While data-driven methods include supervised, unsupervised and semi-supervised approaches, knowledge-based approaches rely on semantic reasoning and modeling [17]. Depending upon the requirements of the application, a combination of two or more approaches belonging to data-driven or knowledge-based classes may be used. These methods are referred to as hybrid approaches.

Technologies such as cloud computing and big data analytics can play an instrumental role in data management and processing using machine learning and advanced data mining techniques. In line with this, Cloud-based body area sensors or CaBAS has emerged as a growing research area that integrates wearable sensors with the MCC paradigm to manage scalability issues associated with provisioning of data-driven pervasive healthcare solutions [15]. Advantages of using this paradigm include improved energy efficiency, creation of annotated data logs, support for event-based processing, development of individual-centric databases and advanced visualization to support self management at the patient level and facilitate decision making at the healthcare provider's level.

## 2.4. Applications

The application layer of the WIoT architecture focuses on provisioning of high quality services to healthcare providers as well as individuals. The user interfaces need to be user-friendly in view of the fact that these solutions may or may not be used by technologically aware individuals. Personalized healthcare finds its best use cases in the elderly who are characteristically not accustomed to using complicated technological interfaces. An example of such interventions include monitoring of tremors in patients suffering from Parkinson's disease using the Smart watch's motion sensor [47].

Conventionally, application was not considered as an independent layer of the IoT architecture for healthcare and was typically integrated with the processing layer. However, with the widening of the application domain for IoT in healthcare and several applications like assistive living and self-management coming into existence, isolating this layer for the sake of application-level novelty became inevitable. The applications of IoT in healthcare [59] are summarized in Table 2. Challenges specific to this layer of the WIoT infrastructure include creation of usable applications that are modeled appropriately to abstract details that are irrelevant and probably not understandable for patients and summarizing details appropriately for healthcare providers to facilitate quick decision making at their end.

Table 2: Healthcare Applications of IoT

| Category | Applications |
|---|---|
| Pervasive Monitoring | Monitoring of patient condition and response to treatment in real time irrespective of patient or clinician's location. |
| Healthcare Management | Management of health records with secure sharing of records between different entities of the healthcare system. |
| | Staff management |
| | Quality assessment in terms of hospital infrastructure and patient outcome and satisfaction |
| | Optimizing resource utilization by managing usage of medication and evaluation the procedures and diagnostics performed. |
| Management of Chronic Patients | Risk Assessment |
| | Tracking, monitoring and quantification of patient's health |
| Medical Research | Monitoring of performance and efficiency of clinical trials |
| | Comparing the effects of treatment and quantifying the patient's functional recovery |
| | Finding new therapeutics |

# 3. Enabling Personalized Healthcare With WIoT

Although, research efforts are far from mature when it comes to application of technologies for development of healthcare-specific applications, the number of applications and possibility of innovation in this domain lies in the infinite space of reality. It would not be wrong to state that we are living in the age of individuality, with all existing systems from e-commerce to daily living solutions turning to personalization for usage and business benefits. Healthcare is not an exception to this unsaid rule.

As shown in Fig. 4, applications of WIoT in personalized healthcare can broadly be divided into five categories namely physical activity monitoring, self management and monitoring, clinical decision support systems for automated diagnosis and treatment, emergency health services and assisted living solutions for elderly and differently-abled. Most wearable devices available today can be connected to the smart phone for monitoring physical activity parameters related to motion, breathing and heart activity. Some of the research projects that focus on this domain include Mobile Sensing Platform [48], Wireless Sensor Data Mining [49] and mHealthDroid [50].

Such applications are specifically useful for patients or individuals suffering from conditions that require physical activity monitoring. One such example is a wearable device and message-based prompting service [60] for monitoring physical activity in individuals who are obese or expected to manifest such a problem. Moreover, integrated system for gymnasiums that monitor physical activity of people who are working out and sets an alarm on detection of a problem also exist [63]. Future research efforts required for evolving physical monitoring devices needs to be centered on clinical integration, privacy, measurement and adherence [61] for it to be widely adopted for clinical management.

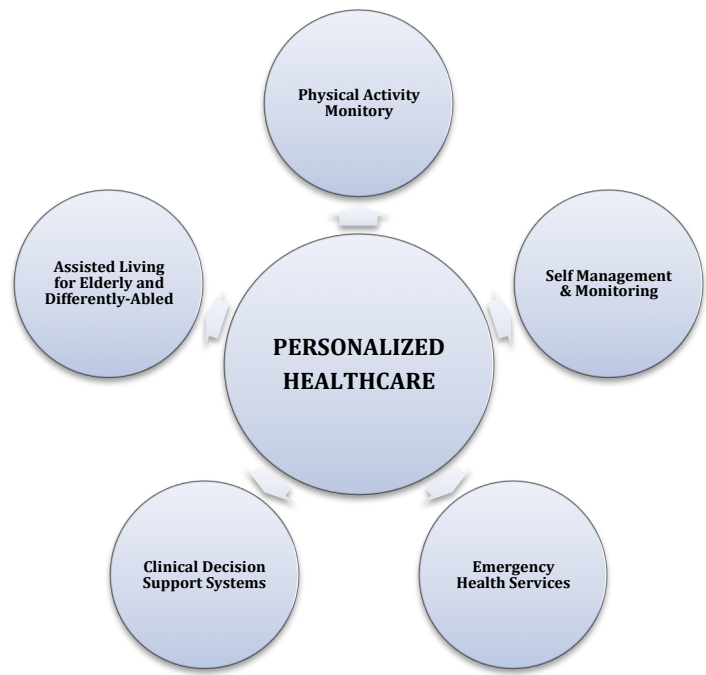

Fig. 4: Personalized Healthcare Applications

Self management and monitoring applications typically concentrate on telemonitoring and aim to improve the individual's quality of life by providing a seamless interface between the caregivers, physicians and individuals. The challenging aspect of these systems is to design solutions that can satisfy the user-specific needs because the success or failure of a therapy largely depends on individual's perception and feedback. Such systems include MOKUS [51] that is designed for arthritis patients' self management and an in-home patient monitoring system [68] for at-home monitoring of patient health.

Clinical decision support systems typically require individual's health data that can be automatically analyzed for prediction of diseases and monitoring the response of the patient to provided treatment. This concept aids treatment planning and facilitates personalized medicine, also bringing down the costs and improving the overall accuracy of the healthcare process. METEOR [52] is a generic infrastructure that captures patient information and aids treatment planning and response monitoring. Besides this, disease-specific solutions like PredictAD [53] also exist. This solution specifically works for the prediction and management of Alzheimer's disease.

Personalized solutions can track individual information to capture movement data and vital parameter measurements for prediction of emergencies. On the onset of such a scenario, automated action from the healthcare providers' end can be initiated. Although, such solutions can be of benefit to all patients suffering from chronic disorders, they are specifically relevant for the elderly population. Emergency Monitoring and Prevention (EMERGE) [54] is a solution that focuses on this aspect of personalized healthcare. Another solution in this category is infant health monitoring system for emergency handling [67].

In continuation to the concept of detection of emergencies and their prevention, individual health data tracking can also be used for development of assisted living solutions for the elderly and differently-abled. Moreover, such solutions can also be expanded to serve patients suffering from chronic and life-threatening diseases so that continuous monitoring can be performed and timely interventions can be made. SMART [55] and PIA [56] are two projects that are working on these facets.

While SMART focuses on care for chronic patients, PIA is a dedicated service for the elderly. Another implemented solution in this domain is AlarmNet [62], which is a monitoring and assisted living solution for residential communities. It provides pervasive healthcare that is adaptable to the varying needs to individuals living in these communities.

In addition to the above mentioned, several applications of WIoT-based personalized healthcare can also be extended to monitoring, management and treatment of emotional issues in individuals. Besides this, hospital and rehabilitation centers' processes [64] can be modeled for development of patient-centric treatment and monitoring applications for use in these domains. The applications of WIoT in personalized healthcare are summarized in Table 3.

Table 3: Applications of WIoT in Personalized Healthcare

| Category | Application | Functionality |
|---|---|---|
| Physical Activity Monitoring | MSP [48] | For this application, the wearable is placed on the waist and is connected to the smart phone for activity monitoring. |
| | WSDN [49] | This platform detects multiple activity using the Android phone carried by the individual and uses supervised learning for detection of the type of activity being performed. |
| | mHealthDroid [50] | This solution connects multiple smart devices together for capturing ambulation and biomedical signals, which are further processed to give alerts and measure parameters like trunk endurance. |
| | SMS-based Notification for Management of Obesity [60] | This application uses SMS-based prompting with Fitbit One for monitoring of physical activity in obese adults. |
| | Sportsman Monitoring | This application monitors vital parameters of a person who is |

| | System [63] | performing workout and sets the alarm in case any abnormal physiological parameter measurement is detected. |
|---|---|---|
| Self Management and Monitoring | MOKUS [51] | Self-disease management for patients suffering from arthritis. |
| | In-Home Health Monitoring System [68] | This solution monitors vital parameters to assess the possibility of deterioration in patients at home. |
| Emergency Health Services | EMERGE [54] | This application performs emergency monitoring and prevention and specifically targets the elderly population. |
| | Infant Health Condition Check [67] | This is a proposed design that measures biometric information of an infant and can be used assess the possibility and cope with an emergency. |
| Clinical Decision Support Systems | METEOR [52] | Generic application centered on planning of treatment and monitoring of response. |
| | PredictAD [53] | This solution specifically works for prediction and management of Alzheimer's disease. |
| Assisted Living for Elderly and Differently-Abled | SMART [55] | This solution is a service for patients suffering from disorders, which include stroke, chronic pain and heart failure. |
| | PIA [56] | This project is an initiative for the elderly who live independently and follows the simple approach that allows elderly people to watch instructional videos uploaded by caregivers for efficient management. |
| | AlarmNet [62] | It is an assisted adaptive solution for residential communities. |
| Miscellaneous | Ubiquitous rehabilitation center [64] | Solution for monitoring of rehabilitation machines. |
| | Etiobe [65] | Application for management of obesity in children |
| | SALSA [66] | Architecture for facilitating response to individual and clinician-specific demands. |

The recent advances in genomics [57] have driven research in the field of personalized medicine with higher zeal and momentum. This is expected to drive a paradigm shift from hospital-centric approach to optimized healthcare services that focus on the treatment and well being of individual subjects. Personalized healthcare is sure to revolutionize this industry and offers innumerable benefits to all entities of the system. However, it also suffers from potential shortcomings, ranging from scientific hurdles to legal and socio-economic challenges that need to be overcome before a transitional shift can be made in one of the most critical sectors of the society.

# 4. Challenges and Future Prospects

There is a rising need for sustainable healthcare for all sections of the society and personalization of treatment and management. In order to achieve the required efficacy, a solid infrastructural foundation needs to be laid for large-scale deployment of wearable sensors and their interaction with conventional medical facilities. The challenges and directions in the integrative use of WIoT for personalized healthcare exist both at the clinical as well as operational levels.

The objective of using technological interventions in the healthcare industry is to allow interaction between patients and healthcare providers beyond the walls of the clinic or hospital. Moreover, there is an ever-insistent demand of healthcare providers from individuals to remain proactive about their medical conditions and overall health. The WIoT infrastructure allows inter-entity communication, thereby allowing immediate feedback on systems, micro-management of patient condition and off-location treatment.

With respect to this facet of WIoT, the level of information to be shared between the patient and healthcare provider can be precise or detailed. For example, an application may just suggest physical activity of 30 minutes to a patient while another application may further fine-grain the information to the type of exercise, along with the details on how to perform them. Similarly, on the clinician's level, the amount of information shared may be a detailed case report or summary. Therefore, deciding on application-specific guidelines and standards is a daunting task because the patient may not be well versed with clinical terms to understand the details and abstraction may be needed. Moreover, the clinician must also just receive the information that is required to make an intervention or suggest an appropriate treatment.

Personalization of treatment is one of the key benefits of WIoT, which is particularly relevant in consideration of the fact that every disease, disorder or syndrome manifests different symptoms in different individuals. Moreover, the intensity of these symptoms shall also vary significantly. This is specifically the case with chronic disorders where doctors face challenges in creating individual treatment plans, as the response of the patient to any prescribed treatment cannot be predicted. Furthermore, the success of a treatment plan depends on the adherence of the plan, which needs to be monitored. In view of these, future research directions in this area include healthcare pattern identification, identification of anomalies and emergency management.

Other issues related with the use of wearable devices include standardization. There are some FDA guidelines that deal with wearable devices used for medical purposes [58]. Since, these devices work on multi-range communication protocols, their safety for human use needs to be established before they can be put to large-scale practice, which requires extensive clinical trials.

Finally, wearable devices may or may not be easy to maintain. On the hardware level, these devices suffer from battery issues. On the software level, one of the biggest challenges faced by designers of wearable devices is providing usable solutions that can present information in an abstractive form with the help of interactive interfaces. Besides this, these solutions must contribute to improving the activity levels of individuals. Furthermore, health data is private and confidential and ensuring safety and privacy and following the legislative guidelines set by different jurisdictions can be an arduous task.

# 5. Conclusion

The synergistic use of Internet of Things (IoT) and wearable technology has led to the development of a new technological paradigm, Wearable Internet of Things (WIoT). This chapter discusses the different aspects of WIoT. The functional components of the WIoT architecture include wearable sensors and mobile devices for sensing, IoT infrastructure for connectivity and cloud-based big data support for data processing. All of these components collectively and progressively contribute to applications like personalized healthcare and assistive living.

This chapter particularly focuses on the use of WIoT for personalization of healthcare services and evidently, the use of this technology can be particularly beneficial for elderly and individuals who are already suffering from a chronic disease. WIoT is capable of revolutionizing healthcare sector by allowing early diagnosis and effective treatment with efficient patient monitoring possible even after the patient has left the hospital. However, in order to achieve success, some inherent challenges like incorporation of healthcare process flows and understanding of standards and requirements need to be tackled.